\documentclass[manuscript]{aastex61}

\usepackage{color}

\received{\today}
\revised{\today}
\accepted{\today}
\submitjournal{ApJ}

\shorttitle{MESAS}
\shortauthors{White et al.}

\begin{document}

\title{MESAS: Measuring the Emission of Stellar Atmospheres at Submm/mm wavelengths}

\correspondingauthor{Jacob White}
\email{jawhite@phas.ubc.ca}

\author[0000-0002-0786-7307]{Jacob Aaron White}
\affil{Department of Physics and Astronomy \\
University of British Columbia \\
6224 Agricultural Rd. \\
Vancouver, BC V6T 1T7, Canada}

\author{Jason Aufdenberg}
\affiliation{Physical Sciences Department\\
Embry-Riddle Aeronautical University\\
600 S Clyde Morris Blvd.\\
Daytona Beach, FL 32114, USA}

\author{A.~C.~Boley}
\affil{Department of Physics and Astronomy \\
University of British Columbia \\
6224 Agricultural Rd. \\
Vancouver, BC V6T 1T7, Canada}

\author{Peter Hauschildt}
\affiliation{Hamburger Sternwarte\\
 Gojenbergsweg 112\\
 21029 Hamburg, Germany}

\author{Meredith Hughes}
\affiliation{Department of Astronomy\\
Van Vleck Observatory\\
Wesleyan University\\
96 Foss Hill Dr.\\
Middletown, CT 06459, USA}

\author{Brenda Matthews}
\affiliation{Herzberg Institute\\
 National Research Council of Canada\\
 5071 W. Saanich Road\\
 Victoria, BC V9E 2E7, Canada}

\author{David Wilner}
\affiliation{Harvard-Smithsonian Center for Astrophysics\\
60 Garden Street\\
Cambridge, MA 02138, USA}

\begin{abstract}

In the early stages of planet formation, small dust grains grow to become mm sized particles in debris disks around stars. These disks can in principle be characterized by their emission at submillimeter and millimeter wavelengths.  Determining both the occurrence and abundance of debris in unresolved circumstellar disks of A-type main-sequence stars requires that the stellar photospheric emission be accurately modeled.  To better constrain the photospheric emission for such systems, we present observations of Sirius A, an A-type star with no known debris, from the JCMT, SMA, and VLA at 0.45, 0.85, 0.88, 1.3, 6.7, and 9.0 mm.  We use these observations to inform a PHOENIX model of Sirius A's atmosphere.  We find the model provides a good match to these data and can be used as a template for the submm/mm emission of other early A-type stars where unresolved debris may be present.  The observations are part of an ongoing observational campaign entitled Measuring the Emission of Stellar Atmospheres at Submm/mm wavelengths (MESAS).

\end{abstract}


\section{Introduction}

Stellar emission at sub-millimeter (submm) to centimeter (cm) wavelengths is nontrivial and is generally not well-constrained \citep{cranmer}. As the Sun is the most well studied star at submm-cm wavelengths, it can be used as an illustrative example for the complexity of stellar emission. In the atmosphere of a ``quiet" Sun, the submm-cm continuum radiation is due primarily to free-free emission \citep{dulk, loukitcheva}. Quiet Sun models predict a 1 mm brightness temperature\footnote{The brightness temperature adopted in this paper is thermal temperature expected for black body emission.} ($\rm T_{B}$) of $\sim 4700$ K, or $\sim80\%$ of the Sun's photosphere $\rm T_{B}$ \citep{wedemeyer}. The ``active" Sun, with strong magnetic fields, is difficult to model because of many contributing emission mechanisms, such as umbral oscillations and explosive events \citep{wang, wedemeyer16}. The $\rm T_{B}$ spectrum of the Sun varies significantly, with a minimum in the far-infrared/submm that is well below the optical $\rm T_{B}$(phot) of 5800 K, followed by a pronounced increase in flux at mm wavelengths, and then a very high $\rm T_{B}$ at cm wavelengths (see blue curves in Fig.\ref{tb_plot}).

The brightness temperature spectrum of the Sun is largely due to its chromosphere and corona. Main sequence A-type stars (as studied here) are thought to not have coronal processes and thus should have a flat mm profile.  However, the details of the brightness temperature spectra are poorly understood, including the expected deviation from the photospheric brightness temperature at a given wavelength. Moreover, the submm-cm emission from main sequence stars other than the Sun is just now being explored, e.g. $\alpha$ Cen \citep{liseau}; AU Mic \citep{macgregor13}; Fomalhaut \citep{white_fom}; Vega \citep{hughes}; Sirius A (this work).

This has direct implications for studying debris disks. \textit{IRAS} and \textit{Spitzer} surveys at 70 and 100 $\mu$m find a high occurrence rate of debris systems around A type stars \citep{su06, thureau}. This however could be an age-selection effect due to A stars being hotter and younger than their F/G/K counterparts \citep{thureau}. Regardless, the high occurrence rate of debris makes A stars common targets in studies that seek to characterize debris. Debris disks are commonly detected through the presence of excess over the expected stellar emission \cite[e.g.,][]{matthews}. This requires an accurate model of the star's spectrum. Therefore, understanding stellar emission is not just important for modeling stellar atmospheric processes, it is also critical for debris disk studies. 

During the early stages of planet formation, small dust grains grow to become mm sized particles, and then eventually form asteroids, comets, and planets \cite[e.g.,][]{johansen,raymond}. After the dispersal of the protoplanetary disk, these minor bodies act as a reservoir of new dust through collisional processes that continuously replenish dust grains from $\mu$m to cm sizes, which would otherwise be cleared on short timescales \citep{matthews}. The presence of heated dust leads to detectable radiation from the grains. Observed emission in excess over stellar emission in the IR to mm can be used as evidence for such debris, and multi-frequency observations can be used to probe grain size distributions through the dust emission's spectral index. However, both the occurrence of submm-cm excess and the dust's spectral index can only be determined with properly calibrated and characterized stellar models that extend to submm-cm wavelengths. Submm-cm observations of debris systems can allow us to infer grain properties, which in turn can be used to constrain material properties of the planetesimals themselves \citep[e.g.,][]{macgregor16}.

Our ability to study the abundance and architecture of extrasolar planetary systems from circumstellar debris relies on an accurate understanding of the stellar emission in these systems. When a debris system is unresolved, such as a distant system or a close-in asteroid belt analogue, it is often difficult to spatially separate the emission of the disk and the emission of the host star. Since observations of debris poor stars are largely non-existent, to approximate the flux contribution from the star itself it is common to extend the far-infrared brightness temperature to longer wavelengths \cite[e.g.,][]{su13}. As can be seen in the case of the Sun, this may over predict the flux at submm wavelengths and under predict the flux at mm-cm wavelengths as the stellar spectrum is not well characterized by a single power law profile (e.g., see the Solar spectrum in Fig.\,\ref{tb_plot}).  A poor estimate of the submm-cm flux from a debris system's host star has significant consequences for the characterization of the disk.

Sirius A is a nearby 225 - 250 Myr A1Vm star \citep{liebert, adelman} with no known debris. At a distance of $2.64\pm0.01$ pc \citep{vanleeuwen}, it is one of our Solar System's closest neighbors and a logical starting point to study the submm-cm stellar emission of A stars. Previous observations of Sirius A at 0.8 - 1.3 mm by \citet{zuckerman} and \citet{chini} find very inconsistent flux values and large uncertainties that vary significantly over the few years between observations. It is also not clear how or if the absolute flux calibration was included in the flux uncertainties. As such, they are not reliable for studying the long wavelength emission of Sirius A.

The data presented here are part of an ongoing observational campaign entitled \textit{Measuring the Emission of Stellar Atmospheres at Submm/mm wavelengths} (MESAS). The MESAS campaign seeks to obtain a broad spectral submm-cm coverage of a range of spectral types in order to build a more complete catalog of stellar emission. In this paper we present JCMT, SMA, and VLA observations of Sirius A. Section 2 describes the details of the observations. Section 3 discusses the modeling procedure to obtain flux densities. Section 4 presents a detailed PHOENIX model of Sirius A's stellar photosphere. Section 5 discusses the implications for circumstellar disks.

\section{Observations}

The flux densities presented here rely on well modeled flux calibrators. Therefore the accuracy of the observed flux of Sirius A in a given observation depends on the absolute flux uncertainty of the flux calibrator. For all stated uncertainties in Table\,1, we include the absolute flux uncertainty of the calibrator and the $\sigma_{RMS}$ of the observations added in quadrature. For JCMT\footnote{In accordance with SCUBA-2 documentation from \citet{dempsey}.}, a 5\% flux calibration uncertainty is used for the 0.85 mm data and a 10\% flux calibration uncertainty is used for the 0.45 mm data. For the SMA\footnote{In accordance with SMA documentation from https://www.cfa.harvard.edu/sma/smaData/} and VLA\footnote{In accordance with VLA documentation from https://science.nrao.edu/.}, a 10\% and 5\% flux calibration uncertainty is used, respectively.

\subsection{JCMT}

The data from the \textit{James Clerk Maxwell Telescope} (JCMT) were acquired on 2017 October 17 and 20 (project ID M17BP008; PI White). The SCUBA-2 instrument \citep{holland} was used which provides continuum observations at 0.45 and 0.85 mm (666 and 353 GHz, respectively). The primary beam is $7.9''$ and $13.0''$ at the two wavelengths, making Sirius A effectively a point source for these observations. The weather conditions were good with a 225 GHz opacity $\tau_{225}$ ranging from 0.032 to 0.033 on the first day of observing and 0.027 to 0.020 on the second day of observing. The total on-source time was 2110 seconds on each day. 

The data were reduced using the {\scriptsize STARLINK} data reduction pipeline {\scriptsize ORAC-DR} \citep{gibb}. From the pipeline, the $REDUCE\_SCAN\_ISOLATED\_SOURCE$ recipe was used for calibration. This recipe is ideal for bright isolated points sources, such as Sirius A. The calibrated images at each wavelength were then co-added together using the {\scriptsize STARLINK} package \textit{PICARD} \citep{gibb_picard} and $MOSAIC\_JCMT\_IMAGES$. The peak flux at the location of Sirius A was taken from the resulting calibrated images (see Table\,1). The $\sigma_{\rm RMS}$ of each map was measured using the \textit{PICARD} package (with the $SCUBA2\_MAPSTATS$ command) and are 1.81 and 17.8 mJy beam$^{-1}$ for the 0.85 and 0.45 mm observations, respectively.

\subsection{SMA}

The data from the \textit{Submillimeter Array} (SMA) were acquired on 2017 January 28 and 2017 March 04 (project ID 2016B-S017; PI White). The observations were requested to be taken approximately a month apart in order to test for short term variability. A third observation was accepted as a filler track and observed on 2018 January 04 (PI Wilner). The observations were centered on Sirius A using J2000 coordinates RA = 06 hr 45 min 08.30 sec and $\delta = -16^{\circ} ~43' ~18.91''$. In order to acquire 0.88 and 1.3 mm (340 and 225 GHz) data simultaneously, the upper side band (USB) of the low frequency receiver was tuned to 345 GHz and the USB of the high frequency receiver was tuned to 230 GHz. The USB of the low frequency tuner used $4\times2.0$ GHz basebands with rest frequency centers at 351.013, 349.000, 347.012, and 345.000 GHz. The lower side band (LSB) used $4\times2.048$ GHz basebands with rest frequency centers at 335.011, 332.998, 331.010, and 328.998 GHz. This gives an effective frequency of 340 GHz (0.88 mm) for the low frequency receiver. The USB of the high frequency receiver used $4\times2.0$ GHz basebands with rest frequency centers at 236.011, 233.998, 232.020, and 229.998 GHz. The spectral windows in the LSB used $4\times2.0$ GHz basebands with rest frequency centers at 220.009, 217.997, 216.009, and 213.996 GHz. This gives an effective frequency of 225 GHz (1.33 mm) for the high frequency receiver. The 2018 January 04 observations only utilized the lower frequency receivers and have an effective frequency of 227 GHz (1.32 mm).

The data were calibrated using Interactive Data Language ({\scriptsize IDL}) with the {\scriptsize MIR} package\footnote{https://www.cfa.harvard.edu/$\sim$cqi/mircook.html}. The 2017 January 28 observations used 7 antennas in the compact configuration with baselines ranging from 9.5 - 68.4 m. There was good weather with a measured 225 GHz opacity of 0.075. Quasar 3C84 was used to calibrate the bandpass, Uranus was used to calibrate the flux, quasars 0725-009 and 0522-364 were used as gain calibrators, and system temperature corrections were applied. The 2017 March 04 observations used 7 antennas in the extended configuration with baselines ranging from 50.0 - 226.0 m. There was good weather with a measured 225 GHz opacity of 0.06. Quasar 3C273 was used to calibrate the bandpass, Ganymede was used to calibrate the flux, quasars 0725-009 and 0522-364 were used as gain calibrators, and system temperature corrections were applied. The 2018 January 04 observations were in the compact configuration. Quasar 3C273 was used to calibrate the bandpass, Uranus was used to calibrate the flux, quasars 0725-009 and 0522-364 were used as gain calibrators, and system temperature corrections were applied.

The SMA data, reduced in {\scriptsize IDL}, are converted to a UV-Fits file format using the IDL task \textit{autofits} and then converted to a {\scriptsize CASA} \citep{casa_reference} MeasurementSet using \textit{MIRFITStoCASA}\footnote{https://www.cfa.harvard.edu/rtdc/SMAdata/process/casa/convertcasa/}. This allows for imaging to be undertaken straightforwardly in the {\scriptsize CASA} environment. The data were imaged with a natural weighting and cleaned using {\scriptsize CASA}'s \textit{CLEAN} algorithm  down to a threshold of $\frac{1}{2}~ \sigma_{\rm RMS}$. The 2017 January 28 observations achieve a sensitivity of $0.85~\rm mJy~beam^{-1}$ and $3.30~\rm mJy~beam^{-1}$ for the 1.3 and 0.88 mm data, respectively. The 2017 March 04 observations achieve a sensitivity of $0.85~\rm mJy~beam^{-1}$ and $1.48~\rm mJy~beam^{-1}$ for the 1.3 and 0.88 mm data, respectively. The 2018 January 04 observations achieve a sensitivity of $0.32~\rm mJy~beam^{-1}$.

\subsection{VLA}

The data from the \textit{Jansky Very Large Array} (VLA) were acquired in Semester 17A on 2017 March 02, 2017 August 21, and 2017 August 22 (project ID 17A-239, PI White). The observations were centered on Sirius A using J2000 coordinates RA = 06 hr 45 min 08.30 sec and $\delta = -16^{\circ} ~43' ~18.91''$. The 2017 March 02 observations were taken in the D antenna configuration with 26 antennas and baselines ranging from 0.035 to 1.03 km. The two 2017 August observations were taken in the C configuration with 28 antennas and baseline ranging from 0.035 to 3.4 km.

The observations were setup to include 6.7 and 9.0 mm (45 and 33 GHz) data in the same Scheduling Block (SB). The 33 GHz data used the Ka Band correlator setup with $4\times2.048$ GHz basebands and rest frequency centers of 28.976, 31.024, 34.976, and 37.024 GHz. This gives an effective frequency of 33 GHz (9.0 mm) for the Ka band. The 45 GHz data used the Q Band correlator setup with $4\times2.048$ GHz basebands and rest frequency centers of 41.024, 43.072, 46.968, and 48.976 GHz. This gives an effective frequency of 45 GHz (6.7 mm) for the Q band. Quasar J0650-1637 was used for bandpass and gain calibration. 3C48 was used as a flux calibration source. Data were reduced using the Common Astronomy Software Applications ({\scriptsize CASA 4.5.0}) pipeline \citep{casa_reference}, which included bandpass, flux, and phase calibrations. 

After consultation with the VLA HelpDesk, the Q band data from the two 2017 August observations had to be discarded due to poor weather and atmospheric conditions adversely effecting the high frequency observations and in particular the reliability of the flux calibrator (3C48). The flux of Sirius A is too low for a reliable self-calibration.

The data were imaged with a natural weighting and cleaned using {\scriptsize CASA}'s \textit{CLEAN} algorithm  down to a threshold of $\frac{1}{2}~ \sigma_{\rm RMS}$. The 2017 March 02 observations achieve a sensitivity of $7~\rm \mu Jy~beam^{-1}$ and $24~\rm \mu Jy~beam^{-1}$ for the 9.0 and 6.7 mm data, respectively. The 2017 August 21 and 22 9.0 mm observations achieve sensitivities of $13~\rm \mu Jy~beam^{-1}$ and $14~\rm \mu Jy~beam^{-1}$.

\section{Visibility Model Fitting}

Sirius A's angular diameter of $0.00602''$, compared to the synthetic beams of the VLA and SMA ($>1''$), make Sirius A effectively a point source. As such, the flux at a given wavelength would in principle be the peak flux density in the reconstructed images. Nevertheless, we choose to take a more robust approach by modeling the visibilites, which is the more standard modeling procedure for interferometic observations. 

We use the {\scriptsize CASA} task \textit{uvmodelfit} to recover the best fit flux density of Sirius A at each wavelength. The task fits a point source to the visibilities of a given data set. A minimum $\chi^{2}$ is converged on through an iterative procedure. The best fit solution is sensitive to the starting parameters, which were taken from the CLEANed images.  Although this algorithm can yield a large reduced $\chi^2$ (as was the case for the SMA 0.88 observations from 2017 January 28), the results are consistent with the peak flux densities per beam as measured directly from the dirty image. Table\,1 summarizes the final values and reduced $\chi^{2}$ for Sirius A at each wavelength. For the JCMT observations, the peak flux and $\sigma_{\rm RMS}$ were taken from the calibrated images. All of the listed uncertainties include absolute flux calibration added in quadrature.

\begin{table}

\centering 
\begin{tabular}{c | c | c | c | c | c | c} 
\hline\ 
   	Wavelength  & Facility & Date & Flux Calibrator & Flux  & Uncertainty & Reduced $\chi^{2}$\\
   		(mm) &  & YYYY MMM DD & & ($\rm mJy$) &  ($\rm m Jy$)& \\
   	\hline \hline
   	0.45 & JCMT & 2017 OCT 17/20 & Uranus & 61.6  & 18.8  & -\\
   	0.85 & JCMT & 2017 OCT 17/20 & Uranus & 15.2  & 1.96  & -\\
	0.88 & SMA  & 2017 JAN 28  & Uranus & 17.4  & 3.74  & 6.72\\
	0.88 & SMA  & 2017 MAR 04  & Ganymede & 15.7  & 2.15  & 1.12\\
	1.3 & SMA   & 2017 JAN 28  & Uranus & 8.52  & 1.21  & 1.91\\
	1.3 & SMA   & 2017 MAR 04  & Ganymede & 6.86  & 0.85  & 1.60\\
	1.3 & SMA   & 2018 JAN 04  & Uranus & 7.49  & 0.87  & 0.977\\
	6.7 & VLA   & 2017 MAR 02  & 3C48 & 0.240 & 0.027 & 1.09\\
	9.0 & VLA   & 2017 MAR 02  & 3C48 & 0.137 & 0.010 & 1.11\\
	9.0 & VLA   & 2017 AUG 21  & 3C48 & 0.119 & 0.014 & 3.85\\
	9.0 & VLA   & 2017 AUG 22  & 3C48 & 0.145 & 0.015 & 7.86\

\label{fit_par}
\end{tabular}
\caption{Summary of flux values for Sirius A. The JCMT flux values were taken as the peak of the emission from the images. The flux uncertainties are the $\sigma_{\rm RMS}$ of the images and the absolute flux calibration uncertainties added in quadrature. For JCMT, a 5\% flux calibration uncertainty is used for the 0.85 mm data and a 10\% flux calibration uncertainty is used for the 0.45 mm data. For the SMA and VLA observations, the flux values were calculated using {\scriptsize CASA}'s \textit{uvmodelfit}. A 10\% flux calibration uncertainty is used for the SMA data and a 5\% flux calibration uncertainty is used for the VLA data.  The uncertainties from \textit{uvmodelfit} are not used for these flux density measurements because they can be underestimated up to a factor of $\sqrt{\chi^{2}_{\rm reduced}}$. Moreover, the uncertainty in the point source flux densities should be comparable to the sensitivity of the observations, motivating the use of $\sigma_{\rm RMS}$. }
\end{table}

\section{PHOENIX Model}

As submm-cm observations of debris poor A-type main sequence stars are largely nonexistent, stellar atmosphere models have not been tested at submm-cm wavelengths. We use the SMA, JCMT, and VLA data presented to test the PHOENIX stellar atmosphere \citep{hauschildt} and compare model photosphere fluxes to our long wavelength observations.

\begin{figure}
\centering
\includegraphics[width=\textwidth]{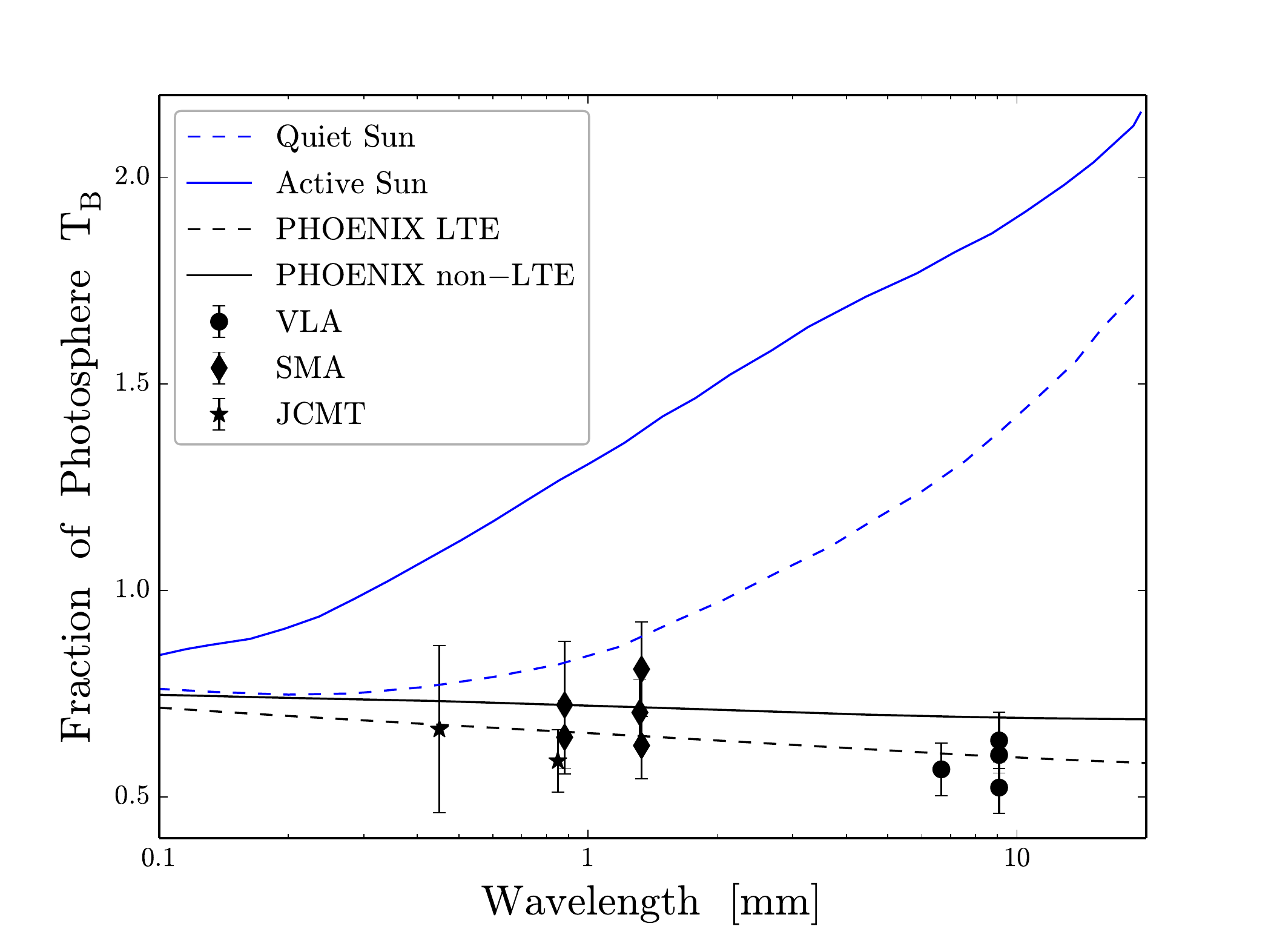} 
\caption{ Submm-cm observations of Sirius A. The two blue curves represent models of the Sun at maximum activity (solid line) and minimum activity (dashed line) from \citet{loukitcheva} and are included for illustrative purposes. The observations of Sirius A are denoted as black stars for the JCMT data, black diamonds for the SMA data, and black circles for the VLA data. The two black curves are PHOENIX models of Sirius A's atmosphere with a non-LTE model (solid line) and a LTE model (dashed line).
\label{tb_plot}
}
\end{figure}

Fig.\,\ref{tb_plot} shows fractional brightness temperature as a function of wavelength. The brightness temperature is normalized to the optical photosphere temperature so that different types of stars can be compared on the same plot. In Fig.\,\ref{tb_plot}, the two blue curves are models of the ``active" and ``quiet" Sun from \cite{loukitcheva} and are included as an illustrative example (normalized to T$_{eff}= 5750$ K). The SMA observations of Sirius A are denoted with black diamonds, the VLA observations are with black circles, and the JCMT observations with black stars.

The two black curves represent spherical 1-D PHOENIX model synthetic spectra of Sirius A's photosphere \citep[version 17.1, see][for an earlier code version]{hauschildt10}. The modeling procedure is similar to \cite{husser}, but now includes the following updates. The atomic line list and the model atoms for the non-local thermodynamic equilibrium (non-LTE) models are based on the 2016 versions of the Kurucz data\footnote{(http://kurucz.harvard.edu/linelists.html)}. For the non-LTE models, new model atoms were generated from the same data so that LTE and non-LTE models are internally consistent with each other. The temperature-pressure structure is computed from the conditions of radiative equilibrium for both LTE and non-LTE models with the assumption of a hydrostatic atmosphere. The non-LTE statistical equations are solved with the 1-D mode of the method discussed in \cite{hauschildt14}, which includes the non-local operator splitting radiative transfer solver and the rate-operator setup as well as the mpack/gmp\footnote{(http://mplapack.sourceforge.net)} based arbitrary precision solver that allows for large dynamic range within the statistical equations.

Fundamental stellar parameters of Sirius A are taken from measured constraints on the angular diameter and bolometric flux \citep{davis} together with the trigonometric parallax and dynamical mass \citep{bond} from its orbit about Sirius B.  For the model, we adopt an effective temperature of T$_{eff}= 9843$ K, a surface gravity of log(g)=4.28, and a radius of 1.71 R$_{\odot}$.  Elemental abundances for 83 elements (complete from H to U except for Tc, Pm, At, Rn, Fr, Ra, and Ac) are taken from recent detailed spectral analyses \citep{landstreet, cowley}, with the exception of F, Ar, and K for which we assume solar abundances.  The model fluxes can be converted to the observed fluxes using the measured limb-darkened angular diameter of 6.04 mas \citep{davis} and plotted in Fig.\ref{tb_plot} by dividing the model fluxes by a Planck function with T$_{eff}= 9843$ K.

The solid black line is the synthetic spectrum from a non-LTE model for Sirius. The model temperature structure and level departure coefficients were converged for 192 atom/ion species (only elements Se, Brm Kr, Sb, I, and Xe have no species in non-LTE) with 37,419 levels in statistical equilibrium and a total of 467,982 spectral lines (bound-bound transitions) in non-LTE and 1,903,022 background ``fuzz" lines from PHOENIX version 17.01.02A.  The dashed line is the synthetic spectrum from an LTE model with the same fundamental parameters and 1,283,018 spectral lines considered in LTE only. The non-LTE temperature structure is flatter and warmer, by 500 K to 1000 K, at the formation depths of the submm-cm continuum relative to the LTE temperature structure.  This temperature structure difference yields a brighter and flatter profile for the non-LTE model compared to LTE model through the submm-cm band.  Neither model appears to be ruled out given present uncertainties in the observed submm-cm fluxes, however the two models diverge further beyond 1 cm and flux measurements at longer wavelengths may more clearly favor one of the models. For a full spectral model of Sirius A's stellar atmosphere, see Aufdenberg et al.\,(in prep).

\section{Implications for Circumstellar Disk Studies}

The observations presented here are the first reliable submm-cm flux densities of Sirius A and A-type main sequence stars in general. To further the detection and characterization of debris disks, it is imperative to understand potential sources of bias in the submm/cm wavelength regime. For unresolved systems, where it is not possible to spatially resolve the stellar and disk emission, an accurate model of the stellar emission is necessary to study the debris. 

In the literature, when estimating the flux contribution of a host star in a debris system, it is common to either assume Blackbody emission at the photosphere temperature or to extrapolate IR observations (also assuming a Blackbody profile) to the observed wavelengths. As is shown in the PHOENIX models of Sirius A's atmosphere, neither  of the assumptions are valid. 

To see how uncertainty in stellar emission can affect the study of debris systems, consider the following example. Fomalhaut is a 200-440 Myr old A3V star \citep{difolco, mamajek} with a well-known extended debris ring at 140 au \citep[e.g.,][]{kalas, boley, white_fom, macgregor}. There is also additional potential IR excess at much closer orbital distances \cite[e.g., an asteroid belt analogue,][]{su13}. However, adopting different stellar models significantly changes the amount of this inferred excess \citep{su16,white_fom}. In particular, ALMA observations at 0.87 and 1.3 mm \citep{boley, su16, white_fom} find a $\rm T_{B}$ of $\sim5500$ K for the central emission (which include both disk and stellar emission). This is much lower than the optical photosphere temperature of $8600\pm200$ K \citep{acke}, and even lower than extrapolating far-IR emission to the mm wavelengths. Confusingly, ATCA 6.6 mm observations find a $\rm T_{B}$ of $> 16000$ K \citep{ricci}, nearly $\sim200\%$ of $\rm T_{B}$(phot), potentially showing a profile similar to the Sun. As Fomalhaut is an A3V star, it is not expected to have a corona, and thus should not have a Solar-like submm-cm emission profile. No conclusions on any unresolved mm debris in the Fomalhaut system can be made until the emission of the host star is known and subtracted from the observed disk flux. 

Fomalhaut and Sirius A are of a similar spectral type and age. Therefore it is reasonable to expect a similar brightness temperature structure for the two stars. The SMA receivers at 0.87 and 1.3 mm (345 and 230 GHz) are complementary to ALMA Bands 7 and 6, which were used to study Fomalhaut. The mm brightness temperature of the central emission in Fomalhaut (which should include the star and the putative inner disk) is consistent with the mm spectra of Sirius A. Taken at face value, the observations and modeling of Sirius A imply that Fomalhaut has no detectable mm debris at close in separations (White et al. in prep.). This would suggest that either the IR observations of Fomalhaut are only detecting stellar emission, that the spectral index of the dust emission is anomalously steep, or that there is only small $\mu$m dust in the inner disk (i.e., more like a Zodiacal Dust analogue than an Asteroid Belt analogue).

The data and modeling presented here assume that there is no detectable variability at any of the observed wavelengths in Sirius A's atmosphere. A-type stars are not expected to have any significant magnetic field activity. However, a weak surface magnetic field of $0.2\pm0.1$ G was detected on Sirius A, although the source of the field is unknown \citep{petit}. The submm-cm spectrum of Sirius A over the $\sim$yr timescale of all the observations presented here is stable enough that if any variability is present, it would either occur only very quickly between observations or over larger timescales than observed. Significant radio variability is however observed in other stars. For example, Herbig Ae/Be B9.5Ve star HD 141569 was found to potentially have significant variability at 9.0 mm \citep{white_141569} as seen with the VLA in 2014 \citep{macgregor16} and 2016 \citep{white_141569}. This variability is prohibiting an accurate characterization of the inner disk around HD 141569. Ongoing observations of Sirius A to determine the long term stability, or variability, are necessary to ensure consistency in the modeling.

Utilizing the full potential of facilities such as ALMA and VLA as well as future facilities such as ngVLA \citep{ngvla} to study debris disks will require increasingly accurate models of the stellar emission of the host stars in those systems.

\section{Summary}

We presented SMA and VLA observations of Sirius A at 0.45, 0.85, 0.88, 1.3, 6.7, and 9.0 mm. We use the observations to compare with both an LTE and non-LTE PHOENIX model atmosphere of Sirius A. The synthetic spectra provides a good match to the data and can be used as a template for stellar emission for A stars at submm-cm wavelengths. An accurate stellar template at long wavelengths is necessary to assess stellar excess due to unresolved features such as a debris disk. The observations are part of an ongoing observational campaign entitled \textit{Measuring the Emission of Stellar Atmospheres at Submm/mm wavelengths} (MESAS).

\acknowledgments

We thank the anonymous referee for the feedback on the manuscript. We thank Mark Gurwell and Chunhua Qi for help with the reduction of the SMA data. J.A.W. and A.C.B. acknowledge support from an NSERC Discovery Grant, the Canadian Foundation for Innovation, The University of British Columbia, and the European Research Council (agreement number 320620). J.P.A wishes to gratefully acknowledge Embry-Riddle Aeronautical University for providing computing time on the Vega supercomputer for model atmosphere computations.

\vspace{5mm}
\facilities{VLA, SMA, JCMT}

\software{{\scriptsize CASA 4.5.0} \citep{casa_reference}; {\scriptsize MIR} (https://www.cfa.harvard.edu/$\sim$cqi/mircook.html); mpack/gmp (http://mplapack.sourceforge.net); PHOENIX version 17.1 \citep{hauschildt10}; {\scriptsize STARLINK} \textit{PICARD} package \citep{gibb_picard} }

\end{document}